\documentclass[aps,twocolumn,superscriptaddress,citeautoscript,nofootinbib,longbibliography]{revtex4-1}
\usepackage{graphicx}
\usepackage{multirow}
\usepackage{xcolor}
\usepackage{times}
\usepackage{amsmath,amsfonts}
\usepackage{bm}
\usepackage{tikz}
\usepackage{dsfont}
\usepackage{mathtools}
\usepackage{setspace}
\usepackage[normalem]{ulem}
\usepackage[caption=false]{subfig}
\usepackage{mnsymbol}
\usepackage{comment}
\usepackage{float}
\usepackage[section]{placeins}

\usepackage[colorlinks=true,linkcolor=magenta,citecolor=magenta]{hyperref}

\def\ba#1\ea{\begin{align}#1\end{align}}
\def\bg#1\eg{\begin{gather}#1\end{gather}}
\def\bpm{\begin{pmatrix}}
\def\epm{\end{pmatrix}}

\newcommand{\tbf}[1]{\textbf{#1}}

\newcommand{\ourtitle}{Correlated normal state fermiology and topological superconductivity in UTe$_2$}

\begin{document}
\title{\ourtitle}

\author{Hong Chul \surname{Choi}}
\thanks {These authors contributed equally to this work.}
\affiliation{Center for Correlated Electron Systems, Institute for Basic Science, Seoul 08826, Korea}
\affiliation{Department of Physics and Astronomy, Seoul National University, Seoul 08826, Korea}

\author{Seung Hun \surname{Lee}}
\thanks {These authors contributed equally to this work.}
\affiliation{Center for Correlated Electron Systems, Institute for Basic Science, Seoul 08826, Korea}
\affiliation{Department of Physics and Astronomy, Seoul National University, Seoul 08826, Korea}
\affiliation{Center for Theoretical Physics (CTP), Seoul National University, Seoul 08826, Korea}

\author{Bohm-Jung Yang}
\email[Electronic address:$~~$]{bjyang@snu.ac.kr}
\affiliation{Center for Correlated Electron Systems, Institute for Basic Science, Seoul 08826, Korea}
\affiliation{Department of Physics and Astronomy, Seoul National University, Seoul 08826, Korea}
\affiliation{Center for Theoretical Physics (CTP), Seoul National University, Seoul 08826, Korea}


\date{\today}
\let\oldaddcontentsline\addcontentsline
\renewcommand{\addcontentsline}[3]{}

\begin{abstract}
UTe$_2$ is a promising candidate for spin-triplet superconductors, in which a paramagnetic normal state becomes superconducting due to spin fluctuations.
The subsequent discovery of various  unusual superconducting properties has promoted the use of UTe$_2$ as an exciting playground to study unconventional superconductivity, but fathoming the normal state fermiology and its influence on the superconductivity still requires further investigation.
Here, we theoretically show that electron correlation induces a dramatic change in the normal state fermiology
with an emergent correlated Fermi surface (FS) driven by Kondo resonance at low temperatures.
This emergent correlated FS can account for various unconventional superconducting properties in a unified way.
In particular, the geometry of the correlated FS can naturally host topological superconductivity 
in the presence of odd-parity pairings, which become the leading instability due to strong ferromagnetic spin fluctuations.
Moreover, two pairs of odd-parity channels appear as accidentally degenerate solutions,
which can naturally explain the multicomponent superconductivity with broken time-reversal symmetry.
Interestingly, the resulting time-reversal breaking superconducting state is a Weyl superconductor in which Weyl points migrate along the correlated FS as the relative magnitude of nearly degenerate pairing solutions varies.
We believe that the correlated normal state fermiology we discovered provides a unified platform to describe the unconventional superconductivity in UTe$_2$.
\end{abstract}

\maketitle

\section{introduction}

Topological superconductors (TSCs) have received significant attention
as a promising platform to achieve stable qubits, using boundary Majorana zero modes
with non-Abelian statistics~\cite{sarma:2015:majorana,sato;2016:majorana,sato:2017:topological}. 
Spin-triplet superconductors are a representative example of TSC candidates~\cite{sato:2010,fu:2010}. 
Since ferromagnetic spin fluctuation is regarded as an origin of  spin-triplet superconductivity~\cite{sigrist:2005:review},
uranium (U)-based compounds in which the coexistence of ferromagnetism and superconductivity was observed~\cite{aoki:2019:review},
have been considered as a promising playground to investigate the physics of spin-triplet superconductivity and related topological properties. 
In this context, the recent discovery of a new U-based superconductor UTe$_2$~\cite{ran:2019:UTe2}, 
for which various pieces of evidence of unconventional spin-triplet superconductivity have been observed,
has immediately come to the forefront of TSC research.
More specifically, UTe$_2$ is expected to be a spin-triplet superconductor, evidenced by the temperature independence of the nuclear magnetic resonance (NMR) Knight shift~\cite{ran:2019:UTe2,nakamine:2019:125Te} and the large upper critical field~\cite{ran:2019:UTe2,aoki:2019:unconventional} above the Pauli limit.
Additionally, follow-up studies have further revealed the unconventional nature of the spin-triplet superconductivity
including  gaplessness, topological properties, time-reversal symmetry (TRS) breaking, and a multicomponent nature ~\cite{sundar:2019,tokunaga:2019:125Te,kittaka:2020,knebel:2019,miyake:2019:metamagnetic,knafo:2019:magnetic,aoki:2020:multiple,lin:2020:MP,braithwaite:2019,thomas:2020,knebel:2020:pressure,ran:2020:pressure,metz:2019:UTe2,bae:2019:anomalous,jiao:2020:nature,hayes:2020,hillier:2012:nonunitary,machida:2001:soc}.

Not only the superconducting state of UTe$_2$ but also its normal state exhibits intriguing characteristics such as heavy fermionic behavior with highly enhanced effective mass~\cite{ran:2019:UTe2} and Kondo resonance~\cite{jiao:2020:nature} arising from strong electron correlation.
Moreover, the normal state of UTe$_2$ is paramagnetic but is under strong magnetic fluctuations without long-range order, contrary to other U-based superconductors with robust ferromagnetism.
This indicates that the fermiology of the correlated paramagnetic normal state under strong spin fluctuations can be the quintessential factor governing the unconventional superconductivity of UTe$_2$, which requires solid theoretical verification

Here we reveal the fermiology of the correlated normal state of UTe$_2$ and the resulting spin-triplet superconductivity with nontrivial topological properties.
Using density functional theory (DFT) plus  dynamical mean field theory (DMFT) calculations with angular momentum-dependent self-energy corrections, we show that the Kondo effect drives the formation of hybridized bands between U $5f$ and conduction electrons, leading to a drastic change in the Fermi surface (FS). 
Namely, at low temperature $T$, we obtain a large correlated FS enclosing the $\Gamma$ point arising from the Kondo effect. 
The emergence of the correlated FS can not only explain the observed heavy fermion physics but also reconcile various types of unconventional superconducting behavior as follows.
First, the geometry of the correlated FS can host topological superconducting phases when odd-parity pairing is developed.  
Second, by solving the linearized Eliashberg equations with the random phase approximation,
we show that odd-parity spin-triplet superconducting channels  become the leading instability due to strong ferromagnetic spin fluctuations.
Moreover, two pairs of odd-parity channels appear as accidentally degenerate solutions, which can naturally explain the multicomponent superconductivity with broken time-reversal symmetry~\cite{hayes:2020}.
Interestingly, we find that the time-reversal breaking superconducting state is a Weyl superconductor in which the positions of Weyl points vary depending on
the relative magnitude of nearly degenerate pairing solutions with the trajectories bounded by the correlated FS.
We believe that the fermiology of the correlated normal state we discovered provides a unified way of understanding the unconventional superconductivity of UTe$_2$.

\section{Band structure calculations}
\begin{figure}[t]
\centering
\includegraphics[width=8.5cm]{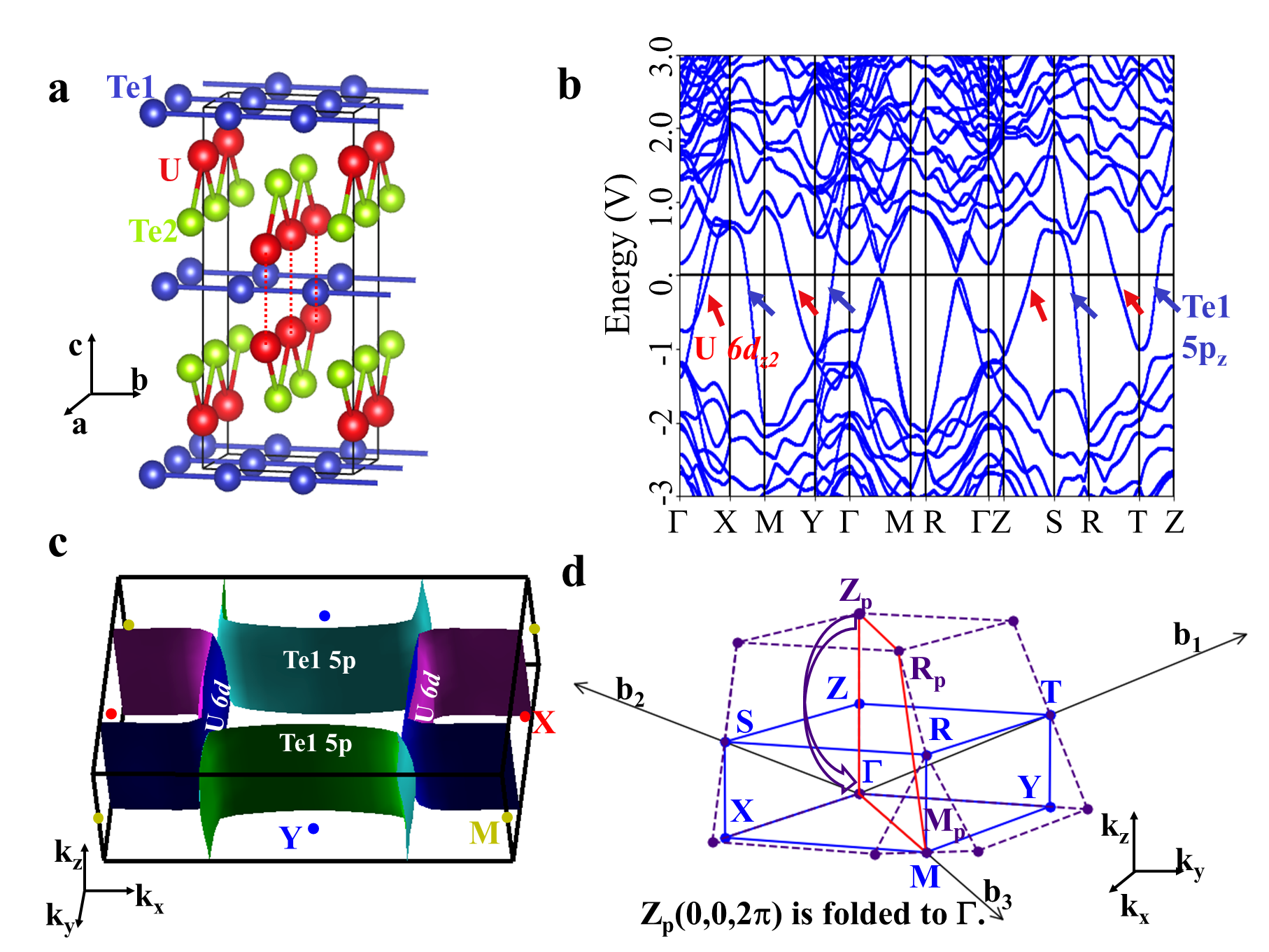}
\caption{\tbf{Crystal structure and open-core DFT calculation results.} \tbf{a,} Schematic crystal structure~\cite{vesta} of UTe$_2$ in the conventional orthorhombic body-centered unit cell. There are two types of Te atoms (Te1 and Te2) as well as U atoms. The blue and red-green solid lines denote the Te1 chains along the $b$-axis and the U-Te2 chains along the $a$-axis, respectively. The red vertical dashed lines indicate the nearest neighbor U atoms along the $c$-axis.
\tbf{b,} Band structure obtained from the open-core DFT calculation with the conventional unit cell without U $5f$ electrons near  $E_F$. The red arrows indicate the hole Fermi pocket around $X$ with a U $5d$ orbital character, and the blue arrows indicate the electron Fermi pocket around $Y$ with a Te1 $5p$ orbital character.
\tbf{c,} FS from the open-core DFT calculations plotted using the conventional unit cell. 
There are two 1D FSs along the $k_x$-axis (U $6d$ states) and the $k_y$ axis (Te1 5p states). 
\tbf{d,} First Brillouin zones (BZs) of the primitive unit cell and the conventional orthorhombic cell. 
Blue large letters indicate the time-reversal invariant momentum (TRIM) in the BZ with the conventional cell.
$Z_p$, $R_p$, and $M_p$ are TRIMs defined in the primitive unit cell. 
$b_1$, $b_2$, and $b_3$ are reciprocal vectors in this primitive unit cell.
The arrow from  $Z_p$ to $\Gamma$ indicates that the $Z_p$ point is folded into the $\Gamma$ point
in the BZ of the conventional unit cell.
}
\label{fig1}
\end{figure}

To date, the band structure of UTe$_2$, with the crystal structure shown in Fig.~\ref{fig1}a, has been reported using various  DFT-based calculations.
For instance, conventional DFT calculations of UTe$_2$ predicted a paramagnetic insulating ground state, 
and thus failed to reproduce its metallic phase at low temperatures.
This occurred because in DFT calculations, the hybridization between U $5f$ states and conduction electrons (U $6d$ and Te $5p$ electrons) is too strong such that a large gap is opened near the Fermi energy ($E_F$)~\cite{ran:2019:UTe2,aoki:2019:unconventional}. 
Introduction of magnetism or an on-site Coulomb interaction ($U$)~\cite{ishizuka:2019:UTe2,ishizuka:2020,shishidou:2020,xu:2019:UTe2,miao:2020:arpes,shick:2019:UTe2,kang2022orbital,UTe2:INS:2020} can partially resolve this issue and restore the metallic ground state. 
However, as the DFT+$U$ method generally suppresses charge fluctuations, the Kondo effect is not properly described.
Therefore, the renormalized FS from the Kondo resonance at low temperatures would be different from a quasi-two-dimensional Fermi surface obtained in previous DFT + $U$ and similar results from some DFT + DMFT  calculations ~\cite{xu:2019:UTe2,miao:2020:arpes}.
Also, in the DFT+DMFT calculations~\cite{kang2022orbital,UTe2:INS:2020,Kang:UTe2optical} performed at intermediate temperatures higher than 10 K, the Kondo effect would not take place effectively.

The electronic structure with localized U $5f$ electrons can be described by using open-core DFT calculations in which U $5f$ electrons are pushed into the core states far  from  $E_F$. 
The resulting band structure in Fig.~\ref{fig1}b supports a quasi-2D FS as shown in Fig.~\ref{fig1}c. 
The origin of this FS is the intrachain and interchain interactions of the two orthogonal quasi-1D atomic chain structures in Fig.~\ref{fig1}a represented by blue solid lines and red-green zigzag lines.
As the Te1 chains lie in the $b$-axis direction (along the $k_x$ axis), Te1 5$p_z$ states appear to be dominant around $E_F$ along the $\Gamma$-$Y$ path. In contrast, the U1-Te2 zigzag chains lie in the $a$-axis direction (along the $k_y$ axis); thus, U 6$d_z$ states  are dominant around  $E_F$ along the $\Gamma$-$X$ path. Such a quasi-2D FS corresponds to the high-temperature phase in which most $5f$ states are localized. 

\begin{figure*}[t]
	\centering
	\includegraphics[width=17 cm]{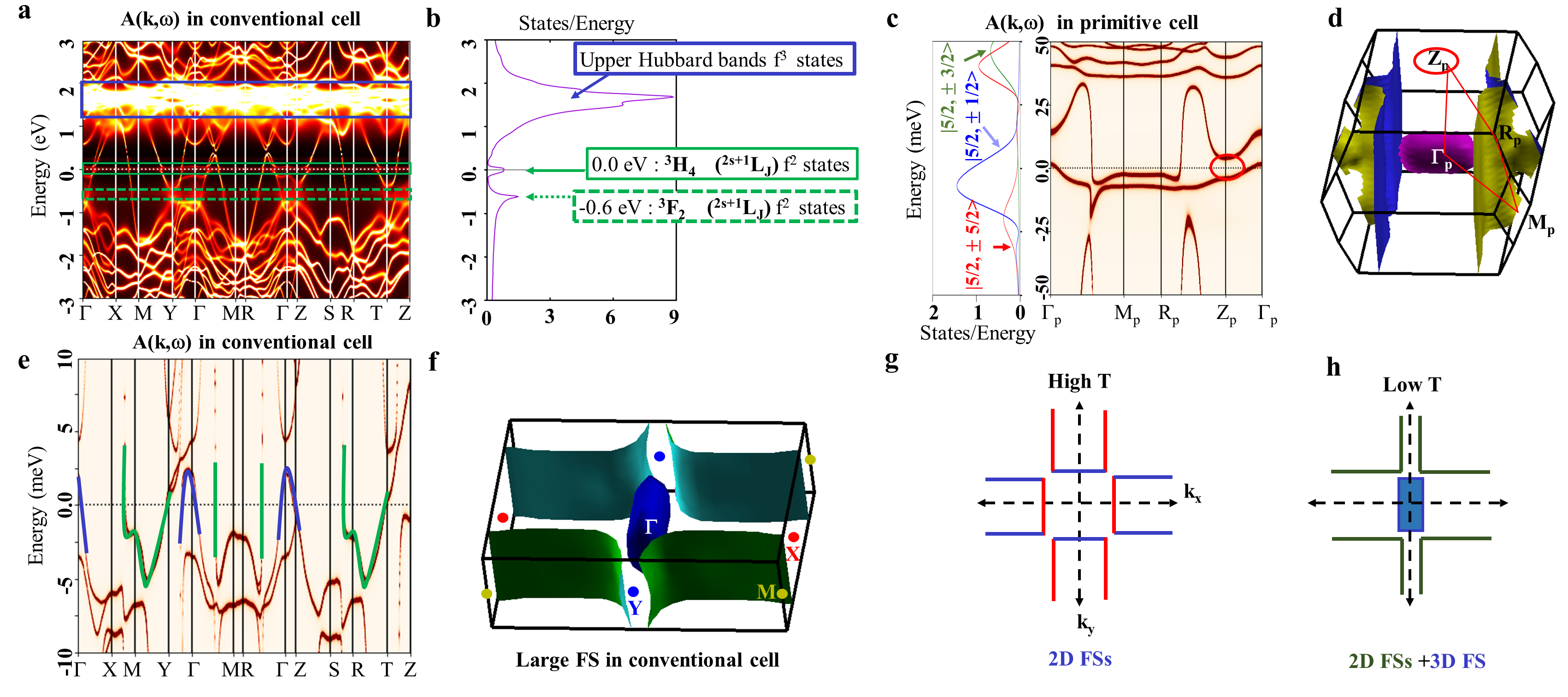}
	\caption{\tbf{DFT+DMFT calculation results.} 
		\tbf{a,} DFT+DMFT spectral functions  
		at $T=11$ K plotted using the BZ of the conventional unit cell.
		\tbf{b,} Emergence of atomic states near  $E_F$ in the DFT+DMFT calculations. The f-electron spectral function is plotted  between -3 eV and 3 eV. The correlation effect drives additional upper Hubbard band ($f^3$) and $f^2$ multiplet states. The atomic multiplet calculation in the DMFT part shows that the $^3 \mathrm{F}_2$ and $^3 \mathrm{H}_4$ of the $f^2$ states  play important roles,  in good agreement with another DFT+DMFT calculation~\cite{miao:2020:arpes}. 
		\tbf{c,} Coherent DFT+DMFT quasiparticle spectrum around $E_F$  
		plotted using the BZ of the primitive unit cell. The $\Gamma_6$ state ($| j=5/2, m_j = \pm 1/2 \rangle$) is dominant around  $E_F$. 
		The red elliptical loop indicates the nearly flat bands near $E_F$ around the $Z_p$ point, 
		which may induce a large spectral weight before the formation of the low-$T$ quasiparticle state.
		\tbf{d,} Large quasiparticle FS in the BZ of the primitive unit cell. 
		The $Z_p$ point is indicated by the red elliptical line.  
		\tbf{e,} Coherent DFT+DMFT quasiparticle spectrum around  $E_F$ plotted using the conventional unit cell. 
		The green- and blue-colored quasiparticle states correspond to the FSs in Fig.~\ref{fig2}f with the same color.
		\tbf{f,} Large quasiparticle FS plotted using the conventional unit cell. 
		The blue hole pocket near the $\Gamma$ point is the correlated 3D FS and the green electron pocket around the $M$ point is a quasi-2D FS, which are indicated by the green and blue lines in Fig.~\ref{fig2}e.
		\tbf{g, h,} Schematic figure describing the Lifshitz transition from the high-T FS to the low-T correlated FS. In the High-$T$ FS, the red and blue lines represent Fermi pockets originating from the U-Te2 and Te1 chains, respectively. In the low-$T$ FS, the green line represents the 2D electron Fermi pocket, and the blue-colored rectangle means the 3D hole Fermi pocket.
	}
	\label{fig2}
\end{figure*}

With decreasing $T$, the emergence of U $5f$ electrons near $E_F$ gives rise to a Lifshitz transition in the FS. 
To capture the Liftshitz transition induced by the Kondo resonance, we perform  DFT+DMFT calculations.
Contrary to the preceding DFT+DMFT calculations, we allow the electronic self-energy to vary depending on the angular momentum $j$ ($j$ = 5/2 and 7/2 for U $5f$ states) and its $z$-component $m_j$ to differentiate the orbital characters.
Throughout our calculations, the spin-orbit coupling is included for all electrons.
The self-energy calculated using the primitive unit cell with two U atoms is obtained at T= 11 K.
The resulting electronic structure of UTe$_2$ at $T=11$ K is shown in Fig.~\ref{fig2}.
In comparison to Fig.~\ref{fig1}b, the spectral function plot shows that  U $5f$ electrons arise around  $E_F$ (inside the green  solid box in Fig.~\ref{fig2}a), but are not yet fully hybridized with conduction electrons at this temperature.
Meanwhile, the upper Hubbard bands of $f^3$ states  are redistributed around 2 eV above  $E_F$ (inside the blue solid box in Fig.~\ref{fig2}a).
Among the $f^2$ multiplets denoted by the symbol $^{2s+1} L_J$, where $s$, $L$, $J$ indicate the spin, orbital, and total angular momentum, respectively, $^3 \mathrm{H}_4$ states play an important role near the $E_F$ (the green solid box), while $^3 \mathrm{F}_2$ states are located -0.6 eV below  $E_F$ (the green dashed box) as shown in Fig.~\ref{fig2}a.
We note that the incoherent spectrum of $^3 \mathrm{F}_2$ states at $-0.6$ eV can be identified with the $-0.6$ eV signal observed in angle-resolved photoemission spectroscopy (ARPES) measurements~\cite{miao:2020:arpes,fujimori:2019} and another DFT+DMFT calculation~\cite{miao:2020:arpes}.
Fig.~\ref{fig2}b summarizes the atomic multiplet distribution in the energy spectrum near  $E_F$.

To obtain the quasiparticle FS from DFT+DMFT, we set the imaginary part of the self-energy to zero~\cite{kang:2021:optical,nomoto:2014:FS,choi:2012:FS}.
The corresponding quasiparticle FSs for the primitive and conventional unit cells are plotted in Figs.~\ref{fig2}d and f, respectively.
The formation of the Kondo resonance state represents the Lifshitz transition to the low-T state with a large FS.
We find that the $\Gamma_{6}$ ($=|j=5/2,m_j=\pm1/2\rangle$) orbital originating from the $^3 \mathrm{H}_4$ state near $E_F$ makes a dominant contribution in the vicinity of $E_F$ from the $m_j$-dependent energy spectrum shown in Fig.~\ref{fig2}c. The enhancement of the $\Gamma_6$ state spectral weight at the Fermi level as decreasing $T$ is consistent with the results in Ref.~\cite{kang2022orbital}.
The corresponding spectral functions in the conventional unit cell show that a hole Fermi pocket encloses the $\Gamma$ point and the electron pocket forms a quasi-2D green Fermi sheet around the $M$ point, as shown in Fig.~\ref{fig2}f, which is equivalent to the FS topology of the primitive cell in Fig.~\ref{fig2}d.
Note that the hole Fermi pocket centered at the $\Gamma$ point is a new discovery of our study, which was absent in previous DFT+DMFT studies that used $m_j$-independent self-energies~\cite{xu:2019:UTe2}, and even in a $m_j$-resolved DFT+DMFT calculation at the temperature above 25K~\cite{kang2022orbital}. This indicates that the appearance of the correlated Fermi surface around the $\Gamma$ point is a novel feature of UTe$_2$ electronic structure that can be captured only by $m_j$-resolved analysis at very low temperatures.

Unlike the DFT+$U$ calculation that achieves the metallic ground state by pushing $5f$ electrons away from  $E_F$, our DFT+DMFT calculation explains how the Kondo effect induces the metallic behavior of UTe$_2$ in which the U $\Gamma_{6}$ state hybridized with Te $p_z$ and U $d_{z^{2}}$ states forms an emergent FS at very low $T$.

Through the Kondo hybridization, the quasi-2D FS surrounding the $X$ or $Y$ point morphs into another quasi-2D FS surrounding the BZ corners and a cylindrical FS closing the $\Gamma$ point [see Fig.~\ref{fig2}g, h].
As $T$ decreases, the cylindrical FS at the BZ center develops a more three-dimensional (3D) character, eventually forming an ellipsoidal closed surface enclosing the $\Gamma$ point.
The 3D nature of the correlated FS is also consistent with the fact that a strong hopping parameter between U atoms along the $c$-axis (see the red dashed lines in Fig.~\ref{fig1}a) comparable to that along the $a$-axis is required to construct a tight-binding model describing the very low-$T$ FS. 
We note that  a strong spectral weight around the $Z$ point of the BZ of the primitive unit cell was measured in a recent high-resolution ARPES study~\cite{miao:2020:arpes}.
We speculate that in the course of the FS evolution from the cylindrical FS to the 3D one, a strong spectral feature can appear from the nearly flat quasiparticle bands around the $Z_p$ point shown in Fig.~\ref{fig2}c, which can be considered a precursor of the formation of a $3D$ correlated FS at very low-$T$ (See the calculation of spectral function in SI).
Additionally, the recently measured nearly isotropic transport property~\cite{eo:2021:iso} further supports the presence of the 3D FS.
 
\section{Superconducting instability} 
\begin{figure}[t]
	\centering
	\includegraphics[width=8.5cm]{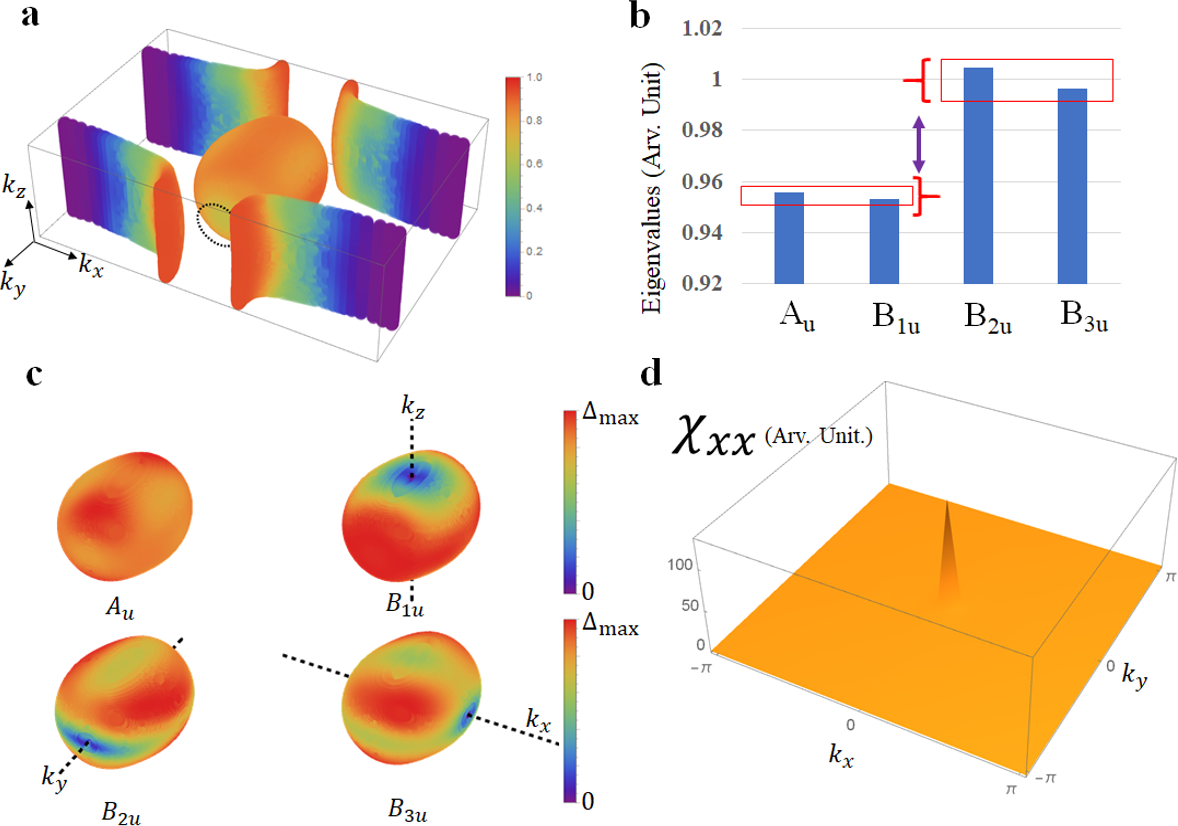}
	\caption{\tbf{Superconductivity from the linearized Eliashberg equation approach.} \tbf{a,} 
		FS reproduced by the tight-binding model. The colors represent the weight of the U $5f$ electron component in the quasiparticle wavefunction on the low-$T$  FS. The region in the black dashed circle shows a relatively low U $5f$ electron contribution on the 3D FS.
		\tbf{b,} Eigenvalues of the linearized Eliashberg equations. The pair of IRs in each red box are almost degenerate.
		\tbf{c,} Schematic superconducting gap structures of four IRs on the correlated 3D FS. 
		\tbf{d,} RPA spin susceptibility in the $a$-axis (easy axis) direction on the $k_xk_y$-plane, which shows strong ferromagnetic fluctuations.
		}
	\label{Fig-gap}
\end{figure}

As the correlated 3D FS consists mainly of U $5f$ electrons with strong local Coulomb interaction $U$ and is thus susceptible to the related spin fluctuations, its emergence at low-$T$ should have a crucial impact on the superconductivity of UTe$_2$.
To explore the superconducting instability of UTe$_2$ by considering the spin fluctuation effect on U $5f$ electrons, we solve the linearized Eliashberg equations given by
\begin{align}
\lambda \Delta_{\xi \xi'}^{\rho} &=-\frac{T}{N}\sum_{k',\xi_{j}} V_{\xi \xi_1 \xi_2 \xi' } (k-k')\nonumber\\
&~~~~~~~~~~~~~~~~~~~~\times G_{\xi_3 \xi_1 }(-k') \Delta_{\xi_3 \xi_4}^{\rho} (k') G_{\xi_4 \xi_2} (k'),
\label{LEl}
\end{align}
where $\xi$, $\xi_1$, $\xi_2$, and $\xi'$ are indices that denote the uranium atomic position and the electron spin, and $k$ and $k'$ indicate the momentum. 
$\Delta^{\rho}$ is a gap function that belongs to an irreducible representation (IR) $\rho$ of the $D_{2h}$ point group. 
For the pairing functions, we use the basis functions listed in Table.~\ref{irrep}.
$V$ is the effective pairing interaction.
As ferromagnetic spin fluctuation is expected to be the origin of the spin-triplet superconductivity in U-based heavy fermion metals~\cite{sigrist:1991:phenomenological}, 
we take into account the spin fluctuation effect from the on-site Coulomb interaction $U$ of $5f$ electrons within the random phase approximation (RPA) (see Fig.~\ref{Fig-gap}d and the SI~\cite{supplement}). 
$G$ is the normal state Green's function of U $5f$ $\Gamma_{6}$ electrons.
To obtain $G$, we construct a tight-binding model that reproduces the correlated normal state fermiology from LDA+DMFT by using U $\Gamma_{6}$, U $d_{z^{2}}$ and Te $p_{z}$ orbitals (Fig.~\ref{Fig-gap}a).  

\begin{table}[t]
	\caption{\tbf{Transformation properties, basis functions, and gap structures of IRs under the $D_{2h}$ point group symmetry.} Here, PN denotes point nodes.}
	\begin{tabular}{|c|c|c|c|c|c|c|c|c|c|c|}
		\hline
		IR & $E$ & $C_{2z}$ & $C_{2y}$ & $C_{2x}$ & $I$ & Basis functions                             	& Gap structure \\ \hline
		$A_{u}$                      & 1   & 1        & 1        & 1        & -1  & $k_{x}\hat{x}$, $k_{y}\hat{y}$, $k_{z}\hat{z}$	& Fully gapped \\ \hline
		$B_{1u}$                     & 1   & 1        & -1       & -1       & -1  & $k_{y}\hat{x}$, $k_{x}\hat{y}$                	& PN ($k_z$-axis) \\ \hline
		$B_{2u}$                     & 1   & -1       & 1        & -1       & -1  & $k_{x}\hat{z}$, $k_{z}\hat{x}$                	& PN ($k_y$-axis) \\ \hline
		$B_{3u}$                     & 1   & -1       & -1       & 1        & -1  & $k_{z}\hat{y}$, $k_{y}\hat{z}$                	& PN ($k_x$-axis) \\ \hline
	\end{tabular}\label{irrep}
\end{table}

By solving the linearized Eliashberg equations, we find that the gap functions belonging to $\rho=A_{u}, B_{1u}, B_{2u}$, and $B_{3u}$ IRs have nonzero eigenvalues (see Fig.~\ref{Fig-gap}b). Interestingly, all four 1D IRs belong to the odd-parity spin-triplet channels, which is consistent with the fact that the strong ferromagnetic fluctuations of U $5f$ electrons give spin-triplet superconductivity. Moreover, the eigenvalues of the four IRs have comparable magnitudes, which is different from previous numerical studies~\cite{ishizuka:2019:UTe2,ishizuka:2020}, in which some of these four IRs were significantly more favored than the others. This might be the direct outcome of the isotropic nature of our correlated 3D FS, which is distinct from the anisotropic cylindrical FS considered in other works.


More specifically, our Eliashberg equation calculations predict two pairs of almost degenerate IRs. Namely, the $A_{u}$ and $B_{1u}$ IRs, and the $B_{2u}$ and $B_{3u}$ IRs appear nearly degenerate, while the $B_{2u}$ and $B_{3u}$ IRs are slightly more favored than the $A_{u}$ and $B_{1u}$ IRs, consistent with a recent renormalization group calculation~\cite{Shaffer2022}.
We note that the IRs of each pair become the same IR when the system's symmetry is lowered by applying an external magnetic field along the $z$-direction.
Among the four IRs, the largest eigenvalue appears in the $B_{2u}$ channel which is favored due to the inhomogeneous distribution of U $5f$ electron wave functions on the $3D$ FS.
That is, as the $3D$ FS has a relatively small U $5f$ electron weight on the $k_y$-axis (see Fig.~\ref{Fig-gap}a), the $B_{2u}$ representation, which has symmetry protected nodes on the $k_y$-axis, has an advantage in lowering the total free energy (see Fig.~\ref{Fig-gap}c and Table~\ref{irrep}).
Notably, the appearance of almost degenerate pairing states is consistent with recent specific heat measurements~\cite{hayes:2020} showing two nearby transition peaks separated by only 80 mK. Since the $B_{2u}$ and $B_{3u}$ IRs have larger eigenvalues than the $A_{u}$ and $B_{1u}$ IRs, we believe the two peaks correspond to the $B_{2u}$ and $B_{3u}$ representations. Moreover, this accidental degeneracy of IRs can lead to a mixed order parameter $B_{2u}+iB_{3u}$, which gives a TRS-breaking Weyl superconductivity (see the SI~\cite{supplement}).

\begin{figure}[t]
	\centering
	\includegraphics[width=8.5cm]{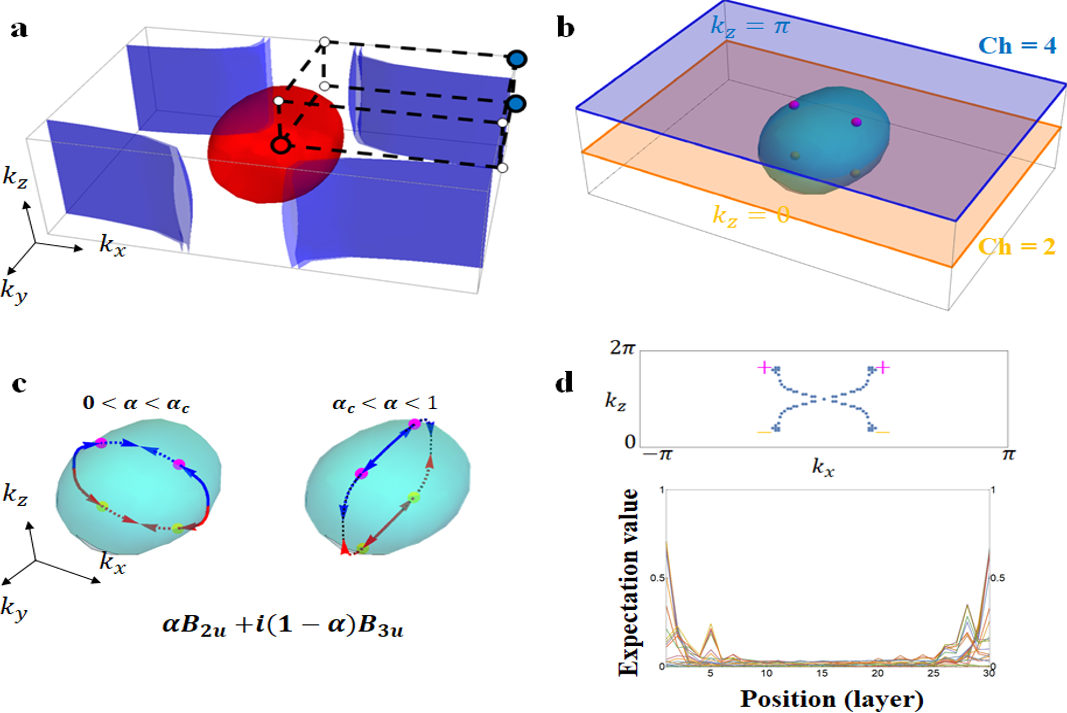}
	\caption{\tbf{Topological superconductivity in UTe$_2$.}
		\tbf{a,} FSs and TRIM points in the BZ of UTe$_2$. 
		\tbf{b,} The Chern numbers carried by the occupied states of the BdG Hamiltonian on the $k_z=0$ plane (orange) and the $k_z=\pi$ plane (blue). The Chern numbers on the two planes differ by 2,  which indicates that the two Weyl nodes (magenta) located between the $k_z=0$ and $k_z=\pi$ planes have the same charge ($+1$ each), while the other two nodes (yellow) located between the $k_z=0$ and $k_z=-\pi$ planes both have charge $-1$.
		\tbf{c,} Trajectory of Weyl nodes. The blue (red) arrows represent the path of nodes with $+$ ($-$) charge. 
		As $\alpha$ increases from 0 to a critical value $\alpha_c$, the Weyl nodes generated by the $\alpha B_{2u}+i(1-\alpha)B_{3u}$ pairing move starting from the $k_x$-axis at $\alpha=0$ to the $k_z$-axis at $\alpha=\alpha_c$. 
		After the nodes with the same charge accidentally meet on the $k_z$-axis at $\alpha=\alpha_c$, 
		they split and move toward the $k_y$-axis  until $\alpha>\alpha_c$ reaches 1.
		\tbf{d,} Zero-energy  Majorana arc states (top), 
		and their position expectation values that show their surface localization (bottom).
	}
	\label{topol}
\end{figure}

\section{Topological superconductivity}

The emergent correlated 3D FS also significantly affects the band topology of the superconducting UTe$_{2}$.
The Bogoliubov-de Gennes (BdG) Hamiltonian of superconducting UTe$_2$ can have various gap structures depending on the symmetry of the pairing function. From symmetry-based analysis, one can show that the $A_{u}$ pairing has a fully gapped spectrum, while the $B_{1u}$, $B_{2u}$, and $B_{3u}$ pairings have nodal points on the $k_z$-, $k_y$-, and $k_x$-axes, respectively,
as summarized in Table~\ref{irrep}.

Let us first consider the fully gapped pairing with the $A_{u}$ IR. 
When TRS is preserved, a $3D$ spinful superconductor such as UTe$_2$ belongs to the Altland-Zirnbauer class DIII whose topological invariant is the $3D$ $\mathbb{Z}$ winding number. In this case, the parity of this $3D$ winding number (a $\mathbb{Z}_2$ invariant) can be captured by counting the numbers of Kramers-degenerate FSs surrounding the TRIM points~\cite{fu:2010,sato:2010,sato:2017:topological}. Since a cylindrical quasi-2D FS always simultaneously encloses a pair of TRIM points, one on the $k_z=0$ plane and the other on the $k_z=\pi$ plane (blue dots in Fig.~\ref{topol}a), it does not contribute to the nontrivial $\mathbb{Z}_2$ invariant. In contrast, the correlated $3D$ FS enclosing only the $\Gamma$ point (the red dot in Fig.~\ref{topol}a) renders the strong bulk $\mathbb{Z}_2$ index nontrivial.
Therefore, UTe$_2$ with the $A_u$ pairing becomes a first-order TSC that hosts gapless Majorana surface states.

In the case of the $B_{1u}$, $B_{2u}$, and $B_{3u}$ pairings, they are guaranteed to have nodal points in the presence of the 3D FS that always intersects with the $k_z$-, $k_y$-, and $k_x$-axes. When TRS is present, these pairing channels support fourfold degenerate gapless points along their high-symmetry lines as summarized in Table 1. 

In contrast, when the nearly degenerate $B_{2u}$ and $B_{3u}$ pairing channels form a complex order parameter in the form of $\alpha B_{2u}+i(1-\alpha)B_{3u}$ ($0\leq\alpha\leq1$), a 4-fold nodal point splits into two 2-fold Weyl points, thus generating a Weyl superconductor. More explicitly, as $\alpha$ becomes slightly larger than zero, two 4-fold nodal points of the $B_{3u}$ pairing state at the intersection between the 3D FS and the $k_x$-axis split into four 2-fold Weyl points located at the intersection between the 3D FS and the $k_x k_z$ plane.
At $\alpha=\alpha_c$, pairs of Weyl points with the same chiral charge merge 
on the $k_z$-axis. The fact that the two Weyl nodes in the $k_z>0$ ($k_z<0$) region have the same monopole charge is further confirmed by computing the Chern number
on the $k_z=0$ and $k_z=\pi$ planes (orange and blue planes in Fig.~\ref{topol}b), which is equal to 2 and 4, respectively.
When $\alpha$ becomes larger than $\alpha_c$, the merged Weyl points with chiral charge $\pm2$ again split into 4 Weyl points, which are located at the intersection between the FS and the $k_y k_z$ plane. Finally, at $\alpha=1$, pairs of Weyl points with opposite chiral charges merge and form 4-fold nodal points of the $B_{2u}$ pairing state on the $k_y$-axis.
In the Weyl superconductor phase, UTe$_2$ hosts  surface Majorana arcs connecting pairs of Weyl nodes with opposite chiral charges projected on the surface BZ as shown in Fig.~\ref{topol}d.
The geometry of the correlated 3D FS promises nontrivial topology of superconducting UTe$_2$, regardless of whether TRS is broken.

\section{Discussion}

In summary, we have performed DFT+DMFT calculations and solved the Eliashberg equations with the tight-binding Hamiltonian 
to study the strongly correlated normal state of UTe$_2$ and the related unconventional spin-triplet superconductivity.
Our correlated electronic structure not only explains the spectral features measured in recent ARPES experiments (See SI for the calculated spectral functions)~\cite{miao:2020:arpes} but also predicts the emergence of correlation-driven $3D$ FS around the $\Gamma$ point at low temperatures,
which can account for various types of anomalous behaviors in both the normal state and the superconducting state~\cite{ishizuka:2019:UTe2,ishizuka:2020}.
In particular, the nearly degenerate spin-triplet solution of the Eliashberg equations naturally predicts a TRS-breaking Weyl superconductor
in which the positions of Weyl points vary depending on the relative magnitude of the nearly degenerate pairing solutions, which can be verified in future experiments. Additionally, the fact that the correlated 3D FS supports the emergence of Weyl points indicates the strongly correlated nature of the Weyl fermions in this system. Thus we propose that UTe$_2$ is a venue to study the intriguing physics of strongly interacting Weyl fermions~\cite{PhysRevB.95.201102,pnas.1715851115}.

Finally, we note that our correlated FS supports the strong ferromagnetic fluctuation, which is consistent with recent experimental results such as the scaling behavior of magnetization~\cite{ran:2019:UTe2,sundar:2019}, anisotropic NMR and dynamical spin susceptibility measurement~\cite{tokunaga:2019:125Te}, and others~\cite{willa:2021,butch:2022,ambika2022possible}. We note that although our correlated normal state fermiology shows that ferromagnetic fluctuation plays a dominant role in the superconducting phase transition of UTe$_2$, there are various recent experimental data that support the importance of incommensurate antiferromagnetic fluctuations including the recent neutron scattering measurements~\cite{UTe2:INS:2020,UTe2:INS:2021-1,UTe2:INS:2021-2}. Resolving the controversy related to the nature of spin fluctuations is definitely one important issue that should be clarified in future research.

The correlated normal state fermiology we obtained shows that ferromagnetic fluctuation plays a dominant role in the superconducting phase transition of UTe$_2$.
We believe that our theory can provide a unified framework to understand the complex behavior of UTe$_2$ and resolve the remaining controversies in this field. 


\vspace{1cm}
\tbf{Method}
\\
\tbf{\emph{Electronic structure calculation}}
The charge self-consistent version of DFT+DMFT~\cite{Kotliar:2006}, as implemented in Ref.~\onlinecite{Haule:2010}, is
based on the full-potential linearized augmented plane-wave (FP-LAPW) band method~\cite{wien2k}.
The correlated $5f$ electrons are treated dynamically by the DMFT local self-energy ($\Sigma(\omega)$), 
while all other delocalized $spd$ electrons are treated on the DFT level.
The charge and spin fluctuations considered in DMFT enable the description of the Kondo effect correctly.
$\Sigma(\omega)$ is calculated from
the corresponding impurity problem, in which  
full atomic interaction matrix is taken into account ($F^0=8.0$ eV,
$F^2= 7.15317919075$ eV,
$F^4= 4.77832369942$ eV,
and $F^6= 3.53367052023$ eV) with $U=8.0$ eV and $J=0.6$ eV~\cite{Cowan}.
A  temperature of 1.1 meV (11 K) is used in the calculations.
To solve the impurity problem, 
we used the continuous quantum Monte Carlo~(CTQMC)~\cite{gull:2011,haule:2007}. 
The  calculated self-energy is analytically continued to the real frequency axis through the maximum entropy method.
The one crossing approximation impurity solver~\cite{Haule:2010} is used to check the validation of the high-temperature CTQMC calculations.
The number of valence electrons from the Fermi surface is calculated using the SKEAF package~\cite{julian2012numerical}.
The Fermi surfaces obtained from the DFT and DFT+DMFT calculations are visualized by the XCrySDen package~\cite{kokalj2003computer}.
\\
\tbf{\emph{Linearized Eliashberg equation}}
The effective pairing potential used to solve the linearized Eliashberg equation reads
\begin{equation}
\hat{V} (q) = -\hat{\Gamma}^{0} \hat{\chi} (q) \hat{\Gamma}^{0} - \hat{\Gamma}^{0},
\end{equation}
where $\hat{\Gamma}^{0}$ is the bare irreducible vertex that describes the on-site Coulomb interaction ($U'$)~\cite{Yanase:2020:LEB}.
We note that $U'$ is defined in the renormalized quasiparticle states near the $E_F$, which is different from $U$ of atomic $5f$ orbitals in the impurity solver in the DMFT loop.
The spin susceptibility $\hat{\chi}(q)$ is calculated from the bare spin susceptibility ($\chi_{0}$) as follows,
\begin{align}
\hat{\chi} (q)  &=  \big[ \hat{1} - \hat{\chi}^{0} (q) \hat{\Gamma}^{0}   \big]^{-1} \hat{\chi}^{0} (q), \\
\end{align}
within the RPA.
\begin{widetext}
To solve the linearized Eliashberg equations
\begin{align}
\lambda \Delta_{\xi \xi'}^{\rho} =-\frac{T}{N}\sum_{k',\xi_{j}} V_{\xi \xi_1 \xi_2 \xi' } (k-k') G_{\xi_3 \xi_1 }(-k') \Delta_{\xi_3 \xi_4}^{\rho} (k') G_{\xi_4 \xi_2} (k'),
\end{align}
we introduce $\phi$ defined as
\begin{align}
[\phi]^{\mu_1 s_1 , \mu_2 s_2}_{\mu_3 s_3 \mu_4 s_4} (\mathbf{k}, i \omega_n =0) & = \sum_{n_1 n_2 } \big[ M_{n_1 n_2 }  \big]^{\mu_1 s_1 , \mu_2 s_2}_{\mu_3 s_3 \mu_4 s_4} \frac{f(\bar{\xi}_{-k,n_1 , \sigma_1 })-f(\xi_{k ,n_2 , \sigma_2})}{\bar{\xi}_{-k,n_1 , \sigma_1 } -\xi_{k ,n_2 , \sigma_2}} \\
&=\sum_{n1} \frac{ \big[ M_{n_1 n_2 }  \big]^{\mu_1 s_1 , \mu_2 s_2}_{\mu_3 s_3 \mu_4 s_4}}{2 \xi_{n1}} \tanh( \frac{\xi}{2T}),
\end{align}
where
\begin{equation}
\big[ M_{n_1 , n_2 } \big]=\big[ u^{\mu_1 s_1}_{n_1 \sigma_1 } (-k) \big]^{*} \big[ u^{\mu_3 s_3}_{n_2 \sigma_3 } (k) \big]^{*}  \big[ u^{\mu_2 s_2}_{n_2 \sigma_2 } (k) \big] \big[ u^{\mu_4 s_4}_{n_1 \sigma_4 } (-k) \big].
\end{equation}
\end{widetext}
Here, $u^{\mu s}_{n \sigma} (k)$ is an eigenstate of the given tight-binding model. $\mu$, $s$, $n$, $\sigma$ are orbital, spin, band index, and pseudospin degrees of freedom, respectively.
\\
\\

\tbf{Data availability}
The data that support the findings of this study are available from the authors upon request.
\\

\tbf{Acknowledgements}
We thank Seokjin Bae and Yun Suk Eo for the fruitful discussions. H.C.C, S.H.L., B.J.Y. were supported by the Institute for Basic Science in Korea (Grant No. IBS-R009-D1). 
S.H.L., B.J.Y. were supported by Samsung  Science and Technology Foundation under Project Number SSTF-BA2002-06,
and the National Research Foundation of Korea (NRF) grant funded by the Korea government (MSIT) (No. 2021R1A2C4002773, and No. NRF-2021R1A5A1032996).
\\

\tbf{Author contributions} BJY initially conceived the project. HCC and SHL contributed to the theoretical analysis and wrote the manuscript with BJY. HCC did all of the ab initio calculations. BJY supervised the project. All authors discussed and commented on the manuscript.
\\

\tbf{Competing financial interest statement} The authors have no competing financial interests to declare.
\\


\let\addcontentsline\oldaddcontentsline
\clearpage
\onecolumngrid
\begin{center}
\bf \large Supplementary Information for ``\ourtitle''
\end{center}
\setcounter{section}{0}
\setcounter{equation}{0}
\setcounter{figure}{0}
\setcounter{table}{0}
\renewcommand{\thesection}{S\arabic{section}}
\renewcommand{\theequation}{S\arabic{equation}}
\renewcommand{\thefigure}{S\arabic{figure}}
\renewcommand{\thetable}{S\arabic{table}}
\hfill \\
\onecolumngrid

\section{Validation of parameters in the DFT+DMFT calculations}

\begin{figure}[t]
\centering
\includegraphics[width=0.95\textwidth]{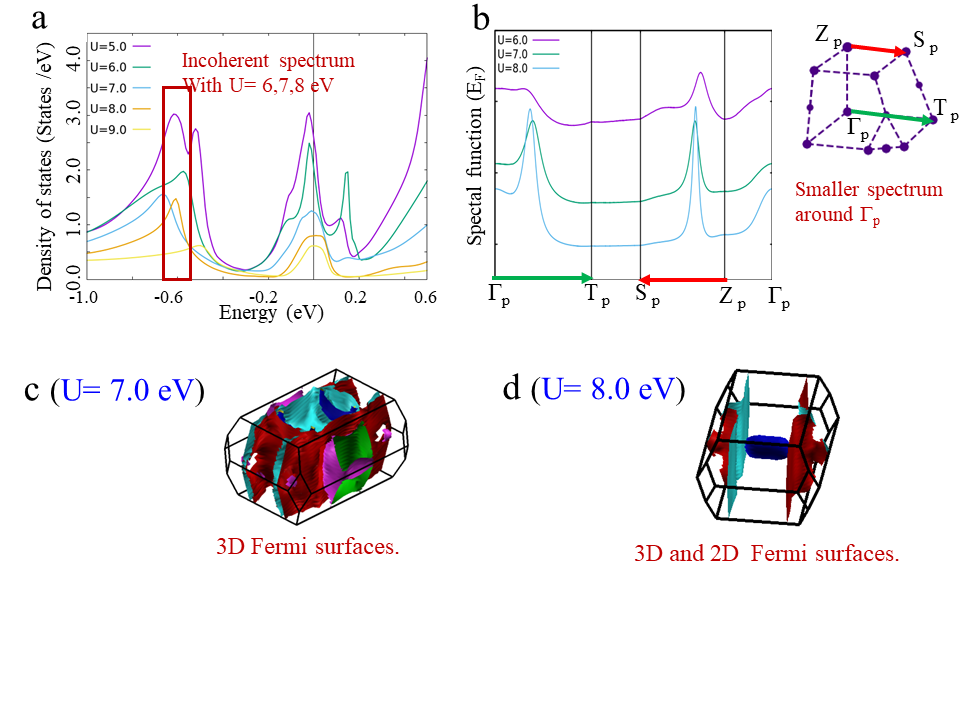}
\caption{\tbf{Validation of the DFT+DMFT calculation.} 
		\tbf{a,} Density of states obtrained by the DFT+DMFT calculations with $U$ = 5-9 eV.
        \tbf{b,} Spectral functions in the momentum space with  the DFT+DMFT calculations with $U$ = 6-8 eV.
        \tbf{c,} Fermi surfaces with  the DFT+DMFT calculations with $U$ = 7 eV.
        \tbf{d,} Fermi surfaces with  the DFT+DMFT calculations with $U$ = 8 eV.
}
\label{SFig-howtoU}
\end{figure}

\begin{figure}[t]
\centering
\includegraphics[width=0.95\textwidth]{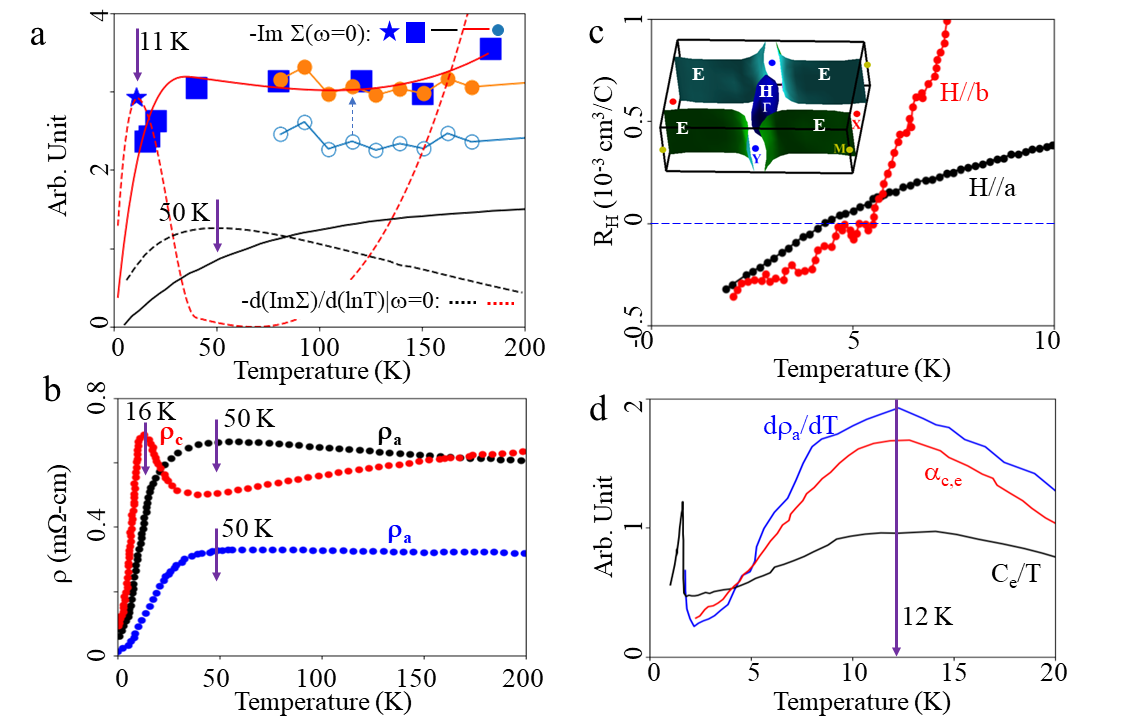}
\caption{\tbf{The computed imaginary part of the self-energy and its comparison to the experiment. } 
		\tbf{a,} The temperature-dependence of the self-energy from the continuous-time quantum Monte Carlo simulation ($U = 8.0$ eV and $J = 0.6$ eV).  The calculated imaginary part ($-\mathrm{Im\Sigma}(\omega=0)$)  of the self-energy of the $\Gamma_6$ state at the Fermi level as a function of temperatures are depicted by the blue star ($T=11$ K) and the blue squares ($T>11$ K). The empty light blue circles mean the calculated $-\mathrm{Im\Sigma}(\omega=0)$ by means of the one-cross approximation ($U =7.0$ eV and $J =0.6$ eV). They are pushed up near the solid red line for a better comparison. The red solid line is the fitted curve from the calculated self-energies with a Fermi liquid behavior at very low temperatures. The red dotted line represents the logarithmic temperature derivative $\frac{\partial\ -\mathrm{Im\Sigma}(\omega=0)}{\partial\log{T}}$ computed from the solid red line. The solid and dotted black lines represent the imaginary part of self-energy and their logarithmic temperature derivatives taken from the previous DFT+DMFT calculation~\cite{xu:2019:UTe2}. For better comparison, red dashed lines and black dashed lines are multiplied by the factors 1.8 and 2.0, respectively. 
        \tbf{b,} The measured resistivity along three crystalline axes as a function of temperature taken from {Y. S. Eo \it{et al.}}~\cite{eo:2021:iso}. The inset shows the very low-temperature Fermi surface. E and H mean electron and hole Fermi pockets, respectively. 
        \tbf{c,} The temperature dependence of Hall coefficient under magnetic field $H= 9$ T taken from {Q. Niu \it{et al.}}~\cite{UTe2:dHvA}. The red ($H//b$) and blue ($H//c$) colors distinguish the direction of the applied magnetic field. The horizontal dashed blue line represents the zero in the $y$-axis. 
        \tbf{d,}The temperature dependence of the magnitude of the electronic contribution to the c-axis thermal expansion (red solid line), Sommerfeld coefficient (black solid line) and temperature derivative of the a-axis resistivity (blue solid line) show a crossover feature at 12 K from {K. Willa \it{et al.}}~\cite{willa:2021}.
}
\label{SFig-Sig}
\end{figure}

\begin{figure}[t]
\centering
\includegraphics[width=0.95\textwidth]{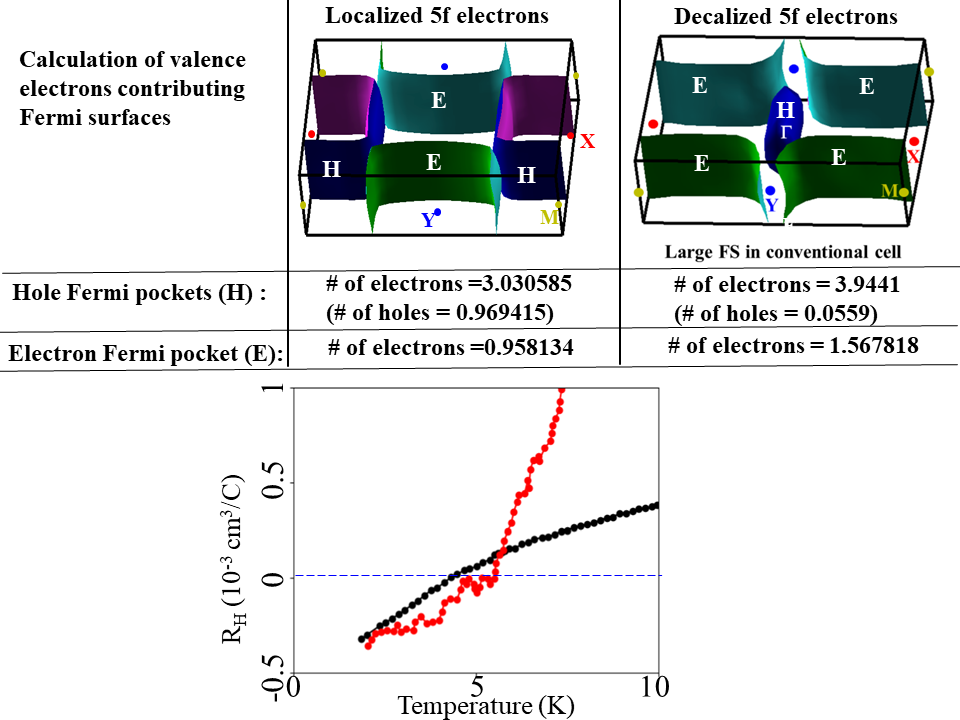}
\caption{\tbf{The Luttinger volume analysis of the DFT+DMFT Fermi surfaces.}	
        Calculation of the number of valence electrons per primitive unit cell in high (localized U $5f$ states) and low temperature (delocalized U $5f$ states) Fermi surfaces. The Hall effect measurements at low temperatures~\cite{UTe2:dHvA} are provided at the bottom.
}
\label{SFig-Luttiner}
\end{figure}

At the beginning of this study, we made lots of efforts to determine the Coulomb interaction used in the CTQMC impurity solver. 
In the previous DFT+DMFT calculations of UTe$_2$~\cite{xu:2019:UTe2,miao:2020:arpes,kang2022orbital,UTe2:INS:2020,Kang:UTe2optical}, the values of $U$ were chosen in between 5 and 8 eV, and the value of $J$ was chosen either 0.57 or 0.6 eV.
We determined the Coulomb interaction by comparing test calculations with different U values in comparison to the ARPES experiment. 
First, we tried to reproduce the incoherent feature around -0.6 eV reported by the ARPES experiments~\cite{fujimori:2019,miao:2020:arpes}.
Fig.~\ref{SFig-howtoU}a shows that uses of $U= 6.0$, 7.0, and 8.0 eV with the fixed $J=0.6$ eV could reproduce such an incoherent feature. 
Then, we excluded $U=6.0$ eV, because the calculation with $U=6.0$ eV showed that spectral weight at $\Gamma_p$ is larger than its neighbor, while the ARPES measurement showed spectral weight at $\Gamma_p$ is smaller than its neighbor~\cite{fujimori:2019}. 
Later, we could confirm that the use of $U=8.0$ eV would be more acceptable than that of $U=7.0$ eV, because the recent experiments showed a 2D-like feature in the geometry of Fermi surfaces observed in the ARPES~\cite{miao:2020:arpes} and the dHvA frequency experiments~\cite{UTe2:dHvA}. 
Fig.~\ref{SFig-howtoU}d shows the DFT+DMFT calculation with $U =8.0$ eV has a 2D Fermi pocket as well as a 3D Fermi surface.

\begin{figure}[t]
\centering
\includegraphics[width=0.95\textwidth]{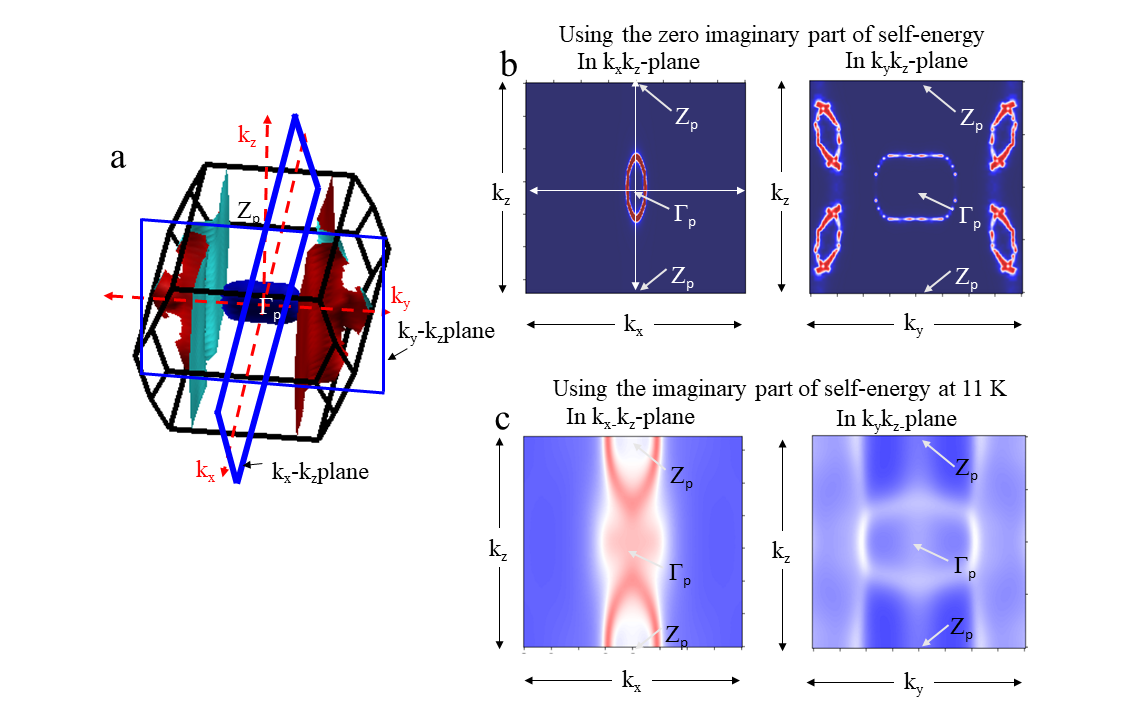}
\caption{\tbf{The enhanced spectral weight around $Z_p$.}	
         \tbf{a,} The Brillouin zone of the primitive body-centered cell in which two planes used in Fig. b and c are specified. $Z_p$ and $\Gamma_p$ are indicated here. 
        \tbf{b,} The spectral functions in the $k_x$$k_z$-plane and the $k_x$$k_z$-plane at very low temperature (when the imaginary part of self-energy is set to be zero). 
        \tbf{c,} The spectral functions in the $k_x$$k_z$-plane and the $k_x$$k_z$-plane at T= 11 K when the self-energy imaginary part is finite. 
}
\label{SFig-AkwZ}
\end{figure}

We note that {Y. Xu \it{et al.}}~\cite{xu:2019:UTe2} analyzed the temperature evolution of the calculated self-energy and its derivative to determine the coherence temperature. Experimentally, a coherence temperature can be identified by the resistivity maximum where the crossover between incoherent and coherent states occurs. The imaginary part of self-energies as a function of temperature and their logarithmic temperature derivatives in {Y. Xu \it{et al.}}~\cite{xu:2019:UTe2} are marked by the black solid and dashed lines, respectively, in Fig.~\ref{SFig-Sig}a. The local maximum of the logarithmic derivative is located at $T=50$ K, which seems to be in good agreement with experimentally measured resistivity maxima along the in-plane axes, depicted by the black and blue dotted lines shown in Fig.~\ref{SFig-Sig}b. 

On the other hand, we find that our calculated temperature-dependent behaviors of the imaginary part of self-energies are different from the previous results. The temperature evolution of the imaginary parts of self-energy at the Fermi level for $\Gamma_6$ states are represented by the blue star (at $T = 11$ K) and blue squares (at $T>11$ K) shown in Fig.~\ref{SFig-Sig}a. To compute the logarithmic derivative, the blue shapes are fitted by the blue solid line assuming the Fermi liquid behavior at very low temperatures. Using this fitted curve, the logarithmic derivative with respect to temperature is also obtained, as indicated by the red dashed line there. There are two important differences between our data and the data in {Y. Xu \it{et al.}}~\cite{xu:2019:UTe2}: (1) the overall magnitudes of the imaginary parts are larger than the previous results (2) our local maximum peak is located at 11 K. To confirm our CTQMC calculations, the one-crossing approximation (OCA) calculation is additionally performed. The results are depicted by the empty light blue circles, which are pushed up and marked by empty orange circles near the red solid line for comparison. The CTQMC and OCA calculations show similar temperature dependences of the imaginary part of the self-energies at higher temperatures. The large imaginary parts at high temperatures are double-checked and confirmed. 

Next, let us discuss how our estimated coherence temperature of 11 K can be compared with the low-temperature experimental measurements. Fig.~\ref{SFig-Sig}b shows the temperature-dependent resistivity along three crystalline axes (blue and black solid lines: in-plane resistivity, red solid line: c-axis resistivity). The c-axis resistivity (red color) develops a peak structure around 16 K while the in-plane resistivities (black and blue colors) show peaks near 50 K. Moreover, as shown in Fig.~\ref{SFig-Sig}d, the thermal expansion coefficient along the a-axis, the specific heat divided by temperature, and the temperature derivative of the a-axis resistivity exhibit a peak structure around 12 K~\cite{willa:2021}.  Therefore, these observations show that 11K would be a more suitable estimation of the coherence temperature than 50 K, because there are several experimental features showing a coherence temperature much lower than 50 K. At this coherence temperature, the heavy quasi-particles start to contribute to the low energy electronic structure and the Fermi surface, which eventually gives the large Fermi surface including the correlated 3D Fermi surface enclosing the $\Gamma_p$ point.

We note that there is more experimental evidence supporting our estimated coherence temperature of 11 K. 
For example, according to the APRES measurement~\cite{miao:2020:arpes}, enhanced spectral weight is started to be developed around the $Z_p$ point at 10 K, which strongly implies a change of the Fermi surface around 10 K. 
In addition,  the sign of the measured Hall coefficient is changed from a positive to a negative value around 5 K~\cite{UTe2:dHvA} (see Fig.~\ref{SFig-Sig}c), which indicates that the dominant carrier has changed from hole-type to electron-type.

According to our result, the Fermi surface without U-$5f$ electrons is composed of a quasi-1D electron Fermi surface and a quasi-1D hole Fermi surface. 
On the other hand, when the U-$5f$ electrons are Kondo-hybridized, the zero-temperature Fermi surface is composed of quasi-two-dimensional electron Fermi surfaces and a three-dimensional hole Fermi surface. 
Here the $3D$ hole Fermi surface is induced by Kondo resonance so that it has a large $f$-electron spectral weight. 
Counting the number of electrons and holes in these two cases ( we computed the Luttinger volume by employing the SKEAF package~\cite{julian2012numerical} as shown in Fig.~\ref{SFig-Luttiner}), we find that the electron carriers become more dominant as the Kondo-resonance is developed. 
Our calculation shows that the number of electrons is increased by almost one electron per unit cell. Simultaneously, the carrier number in the electron pocket of the large Fermi surface is increased by half an electron in comparison to that of the small Fermi surface. The change in the number of valence electrons in the transition from the small Fermi surface to the large Fermi surface reflects the characteristics of a heavy fermion system 
Moreover, as the low-temperature hole Fermi surface induced by U-$5f$ states has a large effective mass, the relevant carrier mobility would be reduced. Taking all these changes into account, one can expect that as the temperature is lowered below the Kondo resonance temperature 11 K, the Hall coefficient should exhibit a sign change from positive to negative, which is consistent with the recent experimental observation (see Fig.~\ref{SFig-Luttiner}). 

Let us note that the observation of the Fermi liquid behaviors (temperature square dependence of the resistivity) for all three crystallographic orientations below 5 K~\cite{eo:2021:iso} further supports the presence of well-defined quasi-particle Fermi surfaces at low temperature.

To sum up, our theoretical estimation of the coherence temperature, 11 K, is compatible with many reported experimental data. Thus, our DFT+DMFT calculation at 11 K could be used to suitably describe the Fermi surface at very low temperatures, which eventually leads to odd-parity spin-triplet superconductivity.

The enhanced spectral weight around the $Z_p$ observed by ARPES was not properly corroborated by the previous DFT+DMFT calculation~\cite{miao:2020:arpes}.
To address this interesting question, we have examined the spectral function at the $Z_p$ point using our DFT+DMFT calculation at $T=11$ K. 
We note that the quasi-particle Fermi surface at very low temperatures in our study was obtained assuming that the imaginary part of the self-energy is zero. This Fermi surface has no weight around the $Z_p$ point as shown in Fig.~\ref{SFig-AkwZ}b. 
Clearly, the 3D hole pocket around the $\Gamma_p$ point can be seen in the 2D Fermi surface plot in the $k_x$$k_z$ and $k_y$$k_z$ planes. 
On the other hand, when we use the self-energy calculated at $T=11$ K keeping its finite imaginary part, the spectral weight of the hole pocket around $\Gamma_p$ becomes faded, and only its broadened frame can be seen.
This means that the 3D Fermi surface cannot be clearly seen around this temperature. 
Interestingly, the same data exhibit some spectral features around the $Z_p$ point as shown in Fig.~\ref{SFig-AkwZ}c. 
Especially, in the $k_x$$k_z$-plane, the strong spectral weight surrounding the $Z_p$ point clearly appears. 
The $k_y$$k_z$-plane shows the slightly enhanced spectral weight at the $Z_p$. point.
Thus, our calculation clearly demonstrates that the enhanced spectral features around the $Z_p$ point can be formed at finite temperature before the formation of the quasi-particle Fermi surface at the $\Gamma_p$ point at zero temperature.

\section{Fermiology}
\begin{figure}[t]
\centering
\includegraphics[width=0.95\textwidth]{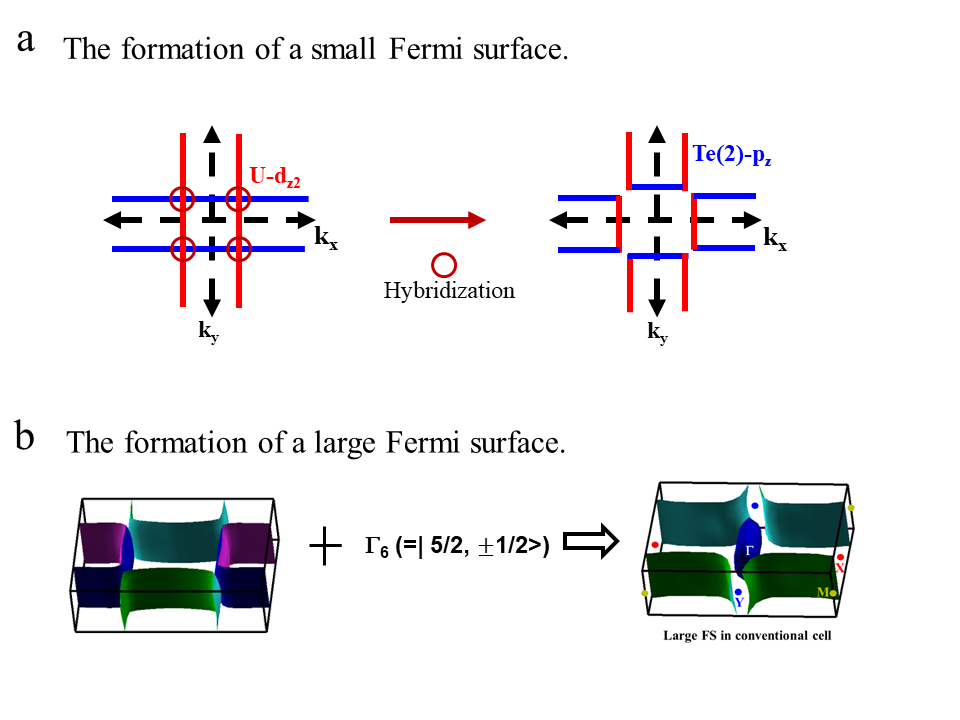}
\caption{\tbf{Fermiology of UTe$_2$.}		
\tbf{a,} The formation of Fermi surfaces in UTe$_2$ at high temperature. The red line presents the 1D Fermi surface due to the zigzag U-Te2 chanins. 
The blue line presents the 1D Fermi surface due to the Te1 chains. The hybridization between them drives a quasi-2D Fermi surface.
\tbf{b,} The formation of Fermi surfaces in UTe$_2$ at low temperature. The Kondo effect between the high-$T$ Fermi surfaces and $\Gamma_6$ quasiparticle states drives a large Fermi surface with the 2D and 3D Fermi pockets. 
}
\label{SFig-FS}
\end{figure}

Fig.~\ref{SFig-FS} explains how the high and low-$T$ FSs are formed. The high-$T$ 2D FS is originating from the hybridization between two 1D FSs. With decreasing $T$, the appearance of the $\Gamma_6$ states near the Fermi level drives a Liftshitz transition from the small FS to the large Fermi surface.

\section{Tight-binding Hamiltonian}

Using U $\Gamma_{6}$, U $d_{z^{2}}$, and Te $p_{z}$ orbitals, the tight-binding model was constructed to reproduce a 3D FS enclosing the $\Gamma$ point, and 2D FS sheets shown in Fig.~\ref{fig2}d,f.

\begin{align}
    \varepsilon_{d}&=1.574276 \\ \nonumber
    \varepsilon_{f} &= -0.17 \\ \nonumber
    \varepsilon_{p} &= -1.9952828 \\ \nonumber
    t^{\textrm{intra}}_{d1}  &=-0.7513 \\ \nonumber
    t^{\textrm{intra}}_{d2}  &=0.0610 \\ \nonumber
    t^{\textrm{intra}}_{f1}  &=0.1 \\ \nonumber
    t^{\textrm{intra}}_{f2}  &=0.01 \\ \nonumber
    t^{\textrm{intra}}_{df1} &=-0.06 \\ \nonumber
    t^{\textrm{intra}}_{df2} &= -0.01 \\ \nonumber
    t^{\textrm{intra}}_{p1}  &=0.1043321 \\ \nonumber
    t^{\textrm{intra}}_{d1}  &=-1.097552 \\ \nonumber
    t^{\textrm{intra}}_{d2}  &=0.095234 \\ \nonumber
    t^{\textrm{inter}}_{f1}  &=-0.1.  \\ \nonumber
    t^{\textrm{inter}}_{f2}  &=0.04  \\ \nonumber
    t^{\textrm{inter}}_{df1} &=-0.25 \\ \nonumber
    t^{\textrm{inter}}_{df2} &= -0.01 \\ \nonumber
    t^{\textrm{inter}}_{p1}  &=-1.970172 \\ \nonumber
    t_{dp}       &=0.25 \\ \nonumber
    t_{fp}       &= 0.13
\end{align}

\begin{figure*}[t]
\centering
\includegraphics[width=17.cm]{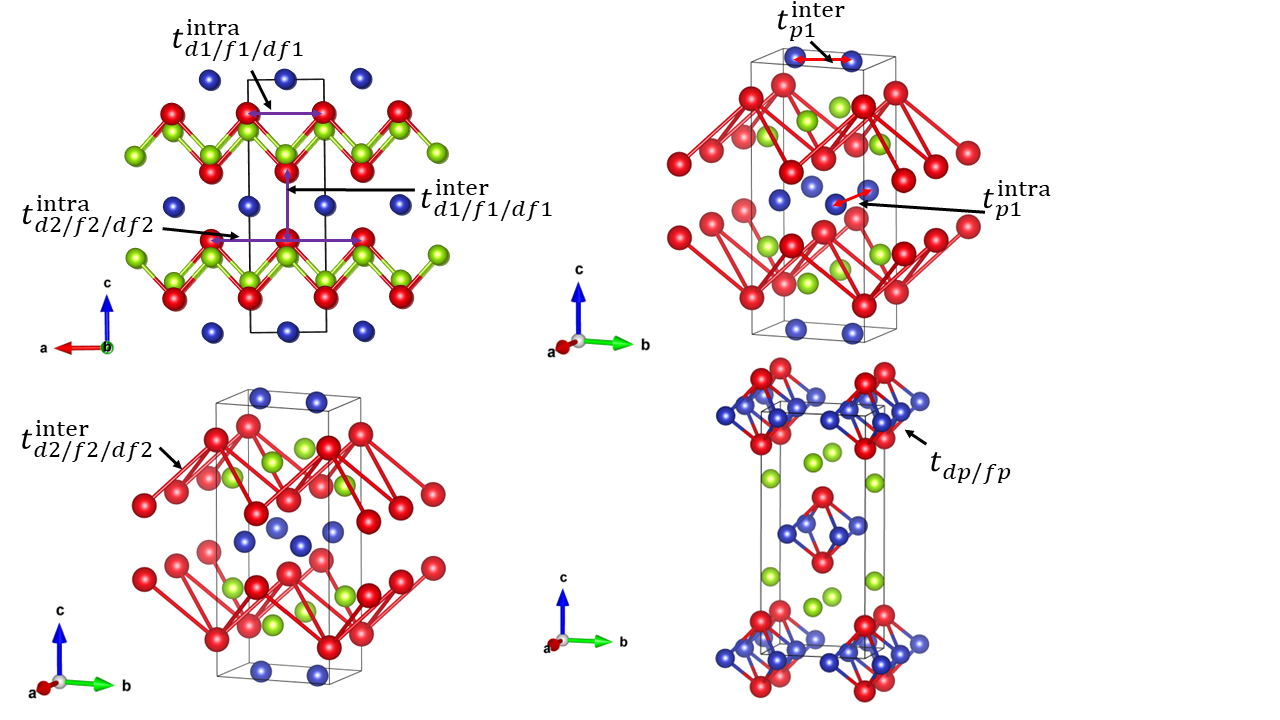}
\caption{\tbf{TB hopping parameters in the conventional orthorhombic unit cell.} 
}
\label{SFig-TB-STR}
\end{figure*}

The above hopping terms are explained in Fig.~\ref{SFig-TB-STR}.

The tight-binding Hamiltonian reads
\begin{align}
    H&= 
    \begin{pmatrix}
    H_{ff} & H_{fd} & H_{fp} \\ 
    H_{df} & H_{dd} & H_{dp} \\ 
    H_{pf} & H_{pd} & H_{pp} \\ 
    \end{pmatrix},
\end{align}
where $H_{ij}$'s are $8\times8$ matrices that represent hopping from $i$-orbital to $j$-orbital.

The below spin-orbit coupling~\cite{ishizuka:2020} with $\alpha_1=0.2$ and $\alpha_2=0.2$ is added to $H_{ff}$ so that the anisotropic spin susceptibility along the $a$-axis is realized.
\begin{align}
    H_{\textrm{ASOC}}&=\big( \alpha_{1} \sin k_y \sigma_{x} -\alpha_{2} \sin k_x \sigma_y \big) \otimes \tau^{\textrm{intra}}_{z} \otimes \tau^{\textrm{inter}}_{0}
\end{align}
where, $\sigma_i$, $\tau^{\textrm{intra}}_j$, and $\tau^{\textrm{inter}}_k$ are the Pauli matrices for the spin, intra-rung sublattice, and inter rung sublattice degrees of freedoms, respectively.

\section{Linearized Eliashberg equation}

\begin{figure}[t]
\centering
\includegraphics[width=0.95\textwidth]{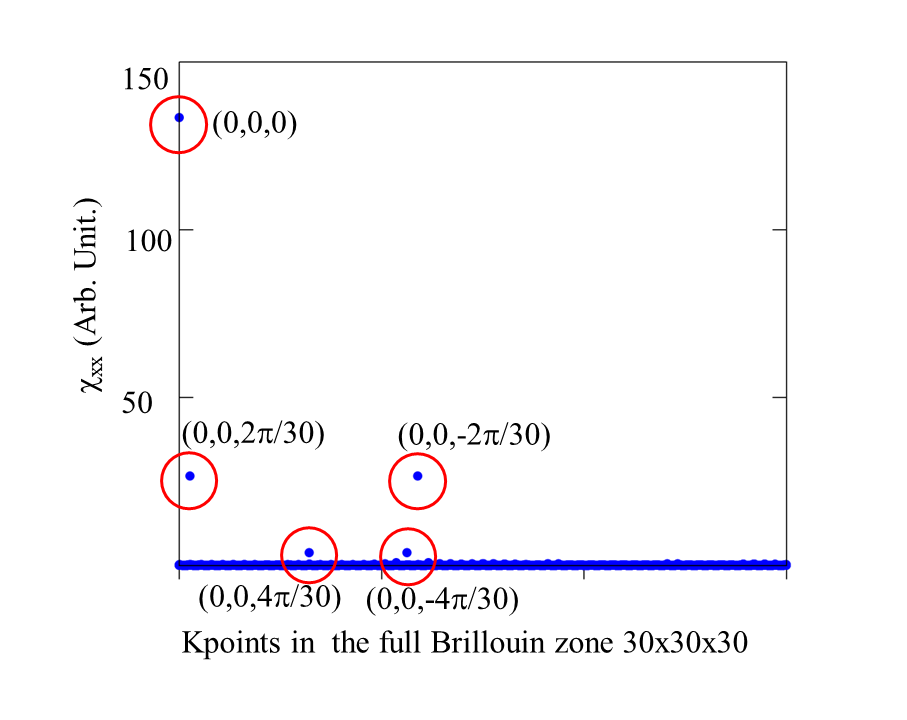}
\caption{\tbf{Spin susceptibility in the full Brillouin zone.} 
The ferromagnetic instability could be seen clearly.  We used a $30\times30\times30$ mesh in the full Brillouin zone. The secondary peaks are located along the $k_z$-axis around the $\Gamma$ point. On-Site Coulomb interaction is set at $U'$= 2.83 eV.
}
\label{SFig-Chi-FBZ}
\end{figure}

\begin{figure}[t]
\centering
\includegraphics[width=0.95\textwidth]{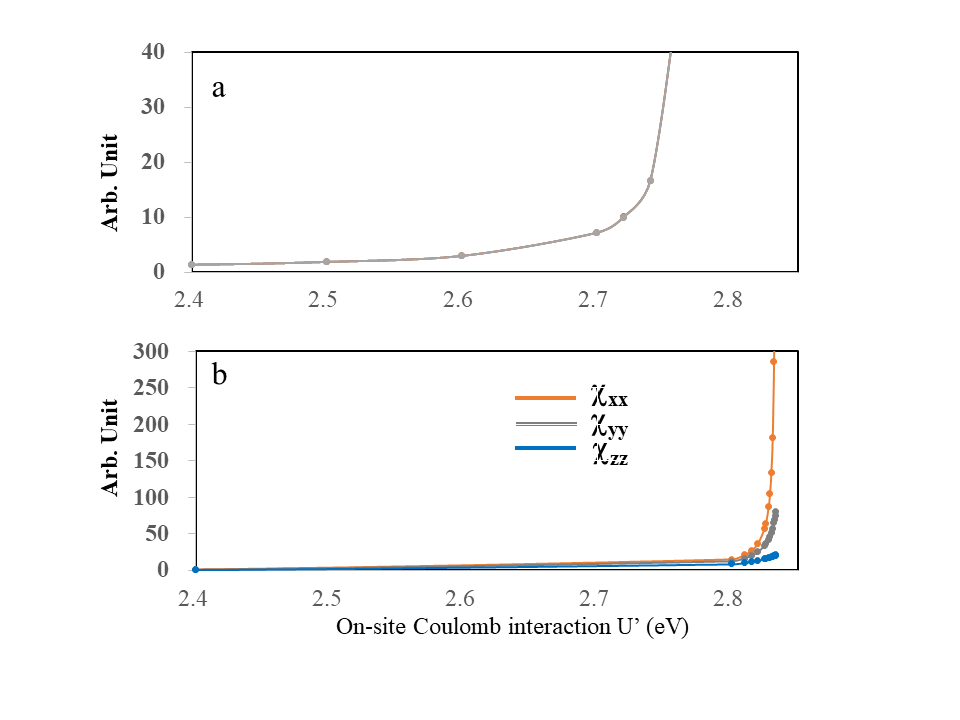}
\caption{\tbf{Spin susceptibility as a function of $U'$ at  $q=\Gamma$  (ferromagnetic fluctuation) with and without spin-orbit coupling (SOC)} 
\tbf{a,} Without the SOC, isotropic ferromagnetic fluctuation occurs. 
\tbf{b,} With the SOC, the easy axis is clearly recognized as the a-axis with the substantial divergence of $\chi_{xx}$.}  

\label{SFig-Chi-FM}
\end{figure}

The effective pairing potential used to solve the linearized Eliashberg equation reads
\begin{equation}
\hat{V} (q) = -\hat{\Gamma}^{0} \hat{\chi} (q) \hat{\Gamma}^{0} - \hat{\Gamma}^{0},
\end{equation}
where $\hat{\Gamma}^{0}$ is the bare irreducible vertex that describes the on-site Coulomb interaction ($U'$)~\cite{Yanase:2020:LEB}.
We note that $U'$ is defined in the renormalized quasiparticle states near the $E_F$, which is different from $U$ of atomic $5f$ orbitals in the impurity solver in the DMFT loop.
The spin susceptibility $\hat{\chi}(q)$ is calculated from the bare spin susceptibility ($\chi_{0}$) as follows,
\begin{align}
\hat{\chi} (q)  &=  \big[ \hat{1} - \hat{\chi}^{0} (q) \hat{\Gamma}^{0}   \big]^{-1} \hat{\chi}^{0} (q), \\
\end{align}
within the RPA.
$\hat{\chi} (q)$ shows a ferromagnetic peak which diverges as $U'$ increases (Fig~\ref{SFig-Chi-FM}).
When spin-orbit coupling is introduced, the $a$-axis is favored as an easy axis.
The ferromagnetic instability was checked through the full Brillouin zone as shown in Fig~\ref{SFig-Chi-FBZ}.

To solve the linearized Eliashberg equations
\begin{align}
\lambda \Delta_{\xi \xi'}^{\rho} =-\frac{T}{N}\sum_{k',\xi_{j}} V_{\xi \xi_1 \xi_2 \xi' } (k-k') G_{\xi_3 \xi_1 }(-k') \Delta_{\xi_3 \xi_4}^{\rho} (k') G_{\xi_4 \xi_2} (k'),
\end{align}
we introduce $\phi$ defined as
\begin{align}
[\phi]^{\mu_1 s_1 , \mu_2 s_2}_{\mu_3 s_3 \mu_4 s_4} (\mathbf{k}, i \omega_n =0) & = \sum_{n_1 n_2 } \big[ M_{n_1 n_2 }  \big]^{\mu_1 s_1 , \mu_2 s_2}_{\mu_3 s_3 \mu_4 s_4} \frac{f(\bar{\xi}_{-k,n_1 , \sigma_1 })-f(\xi_{k ,n_2 , \sigma_2})}{\bar{\xi}_{-k,n_1 , \sigma_1 } -\xi_{k ,n_2 , \sigma_2}} \\
&=\sum_{n1} \frac{ \big[ M_{n_1 n_2 }  \big]^{\mu_1 s_1 , \mu_2 s_2}_{\mu_3 s_3 \mu_4 s_4}}{2 \xi_{n1}} \tanh( \frac{\xi}{2T}),
\end{align}
where
\begin{equation}
\big[ M_{n_1 , n_2 } \big]=\big[ u^{\mu_1 s_1}_{n_1 \sigma_1 } (-k) \big]^{*} \big[ u^{\mu_3 s_3}_{n_2 \sigma_3 } (k) \big]^{*}  \big[ u^{\mu_2 s_2}_{n_2 \sigma_2 } (k) \big] \big[ u^{\mu_4 s_4}_{n_1 \sigma_4 } (-k) \big].
\end{equation}

\begin{figure}[t]
\centering
\includegraphics[width=0.95\textwidth]{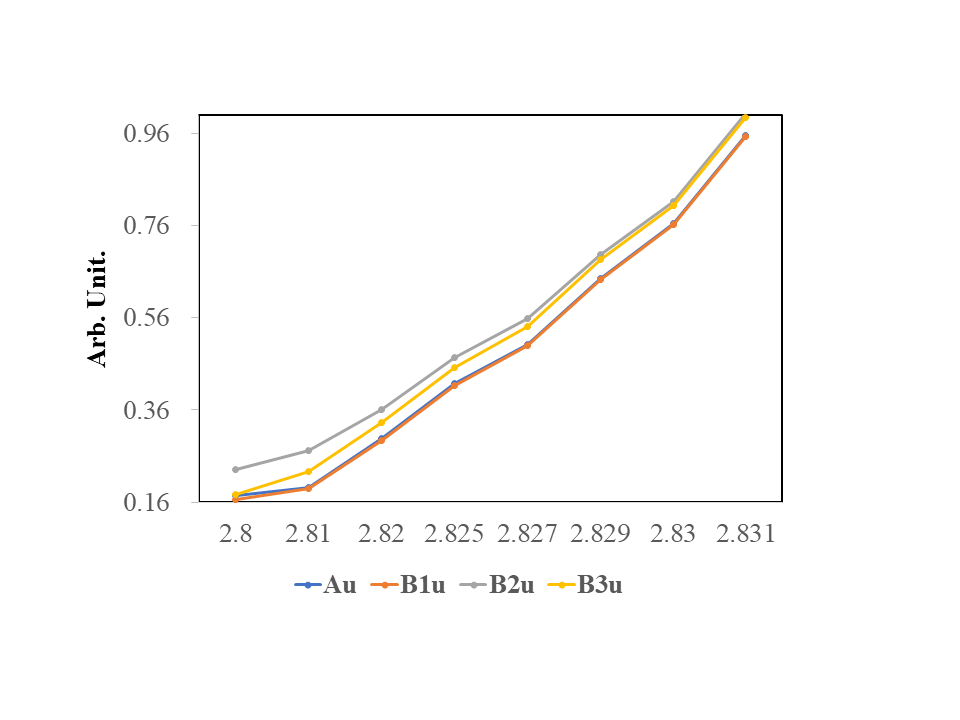}
\caption{\tbf{Linearized Eliashberg equation as a function of $U'$.}  Eigenvalues are plotted as a function of $U'$. 
}
\label{SFig-LEE}
\end{figure}

Here, $u^{\mu s}_{n \sigma} (k)$ is an eigenstate of the given tight-binding model. $\mu$, $s$, $n$, and $\sigma$ are orbital, spin, band index, and pseudospin degrees of freedom, respectively.
Solving the linearized Eliashberg equations at $T=0.003$ eV, we obtain eigenvalues as a function of U' shown in Fig.~\ref{SFig-LEE}.
As U' is increased, B$_{2u}$ and B$_{3u}$ pairings become degenerated and Their eigenvalues is larger than the other degenerated A$_{u}$ and B$_{1u}$ pairings.

\section{Topological invariant}

Under TRS and spatial inversion symmetry, the (strong) topological invariant ($\omega$) in an odd-parity superconductor can be defined as below, 
\begin{align}
    \omega = \frac{1}{2} \sum_{K_i} n(K_i)~~(mod~2),
\end{align}
where $\nu_1$, $\nu_2$ and $\nu_3$ are the weak invariants defined on the $kx=0$, $k_y =0$, and $k_z =0$ planes.
The topological invariants for Fig.~\ref{fig1}c, Fig.~\ref{fig2}f are summarized in Table~\ref{TI-table}.
 \begin{table}
	\caption{\tbf{Topological invariants in $D_{2h}$.}}
    \begin{tabular}{ | l |  l | l | l | l |l |l |l |l |l |}
    \hline
     case & $\Gamma$ & $X$ & $M$ & $Y$ & $Z$ & $S$ & $T$ & $R$ & ($\omega$,$\nu_1$,$\nu_1$,$\nu_1$) \\ \hline
     Open-core & 0 & 4 & 0 & -4 & 0 & 4 & -4 & 0 & (0,0,0,0) \\ \hline
     DFT+DMFT  & 2 & 4 & 8 &  4 & 4 & 4 &  4 & 8 & (1,1,1,1) \\ \hline
    \end{tabular}
   \label{TI-table}
\end{table}

\begin{figure}[t]
	\centering
	\includegraphics[width=0.95\textwidth]{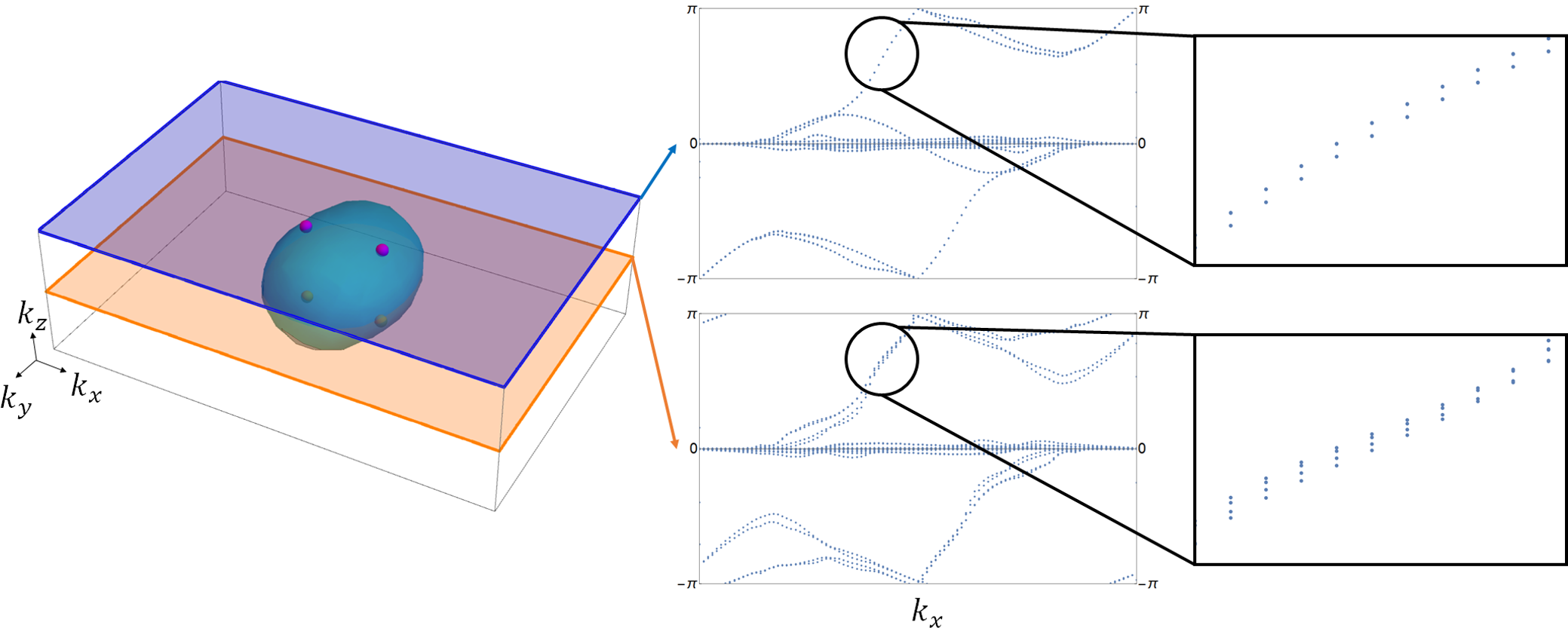}
	\caption{\tbf{The Wilson loop eigenvalue spectra}
		 The Wilson loop eigenvalue spectra of the occupied states of the BdG Hamiltonian of UTe$_2$ with arbitrary pairing function in $B_{2u}+iB_{3u}$ symmetry on the $k_z=0$ plane (orange) and the $k_z=\pi$ plane (blue). The winding numbers of the two Wilson loops show that the Chern numbers carried by the two planes differ by 2. It indicates that two Weyl nodes located between the $k_z=0$ and $k_z=\pi$ planes have the same charge.
	}
	\label{UTe2-SFIg-Chern}
\end{figure}

When TRS is broken by $B_{2u}+iB_{3u}$ pairing, the 2D BdG Hamiltonians on the $k_z=0$ plane and the $k_z=\pi$ plane are fully gapped. Thus we can calculate the Chern number carried by the occupied bands for each plane. The difference between the Chern numbers of the two planes is equal to the total monopole charge of Weyl nodes located between the two planes. As shown in Fig.~\ref{UTe2-SFIg-Chern}, the two Weyl nodes in $0<k_z<\pi$ region has $+1$ charge for each, and the other two in $-\pi<k_z<0$ region has $-1$ charge for each.

\section{Spontaneous TRS breaking}

We consider the most general form of Ginzburg-Landau free energy density that describes a system with two almost degenerate complex order parameters $\eta_1$ and $\eta_2$. The free energy density reads
\begin{align}
\label{freeenergy}
\mathcal{F}=&\frac{r_0}{2}(\eta_1^*\eta_1+\eta_2^*\eta_2)-\frac{x}{2}(\eta_1^*\eta_1-\eta_2^*\eta_2)+\frac{u_+}{8}(\eta_1^*\eta_1+\eta_2^*\eta_2)^2\\ \nonumber
&+\frac{u_-}{8}(\eta_1^*\eta_1-\eta_2^*\eta_2)^2-\frac{g}{8}[-i(\eta_1^*\eta_2-\eta_1\eta_2^*)]^2,
\end{align}
where $u$, $g$, $\lambda$ are phenomenological coupling coefficients and $u\pm=u\pm(g+\lambda)$.
Let $\eta_1=r_1e^i\theta_1$ and $\eta_2=r_2e^i\theta_2$. By differentiating Eq.~\ref{freeenergy} by $r_1$, $r_2$, $\theta_1$, and $\theta_2$, we obtain the extremum conditions for $\mathcal{F}$. Particularly, when differentiated by $\theta_1$ or $\theta_2$, the last term in Eq.~\ref{freeenergy} gives $\sin{2(\theta_2-\theta_1)}=0$. Thus, when $u_\pm>0$ and $g>0$, $\mathcal{F}$ is minimized if $\theta_2-\theta_1=n\pi/2$ where $n$ is an arbitrary integer. In such a case, $\eta_1+i\eta_2$ becomes the ground state order parameter of this system.




\bibliography{mybiblio.bib}

\end{document}